\documentclass[twocolumn,amsmath,amssymb,floatfix]{revtex4} \usepackage{bm}
\usepackage{amsmath}
\usepackage{epsf}
\input epsf
\newcommand{\el}{l}
\newcommand{\bn}{\hat{\bf n}}

\newcommand{\bl}{{\bf l}}

\newcommand{\sYlm}[2]{\, {\vphantom{Y}}_{#1}Y_{\el_#2 m_#2}}
\newcommand{\wj}[6]{\left(
                           \begin{array}{ccc}
        \! #1\! & #2\!  & #3\!  \\
        \! #4\! & #5\!  & #6\!
                           \end{array}
                   \right)}

\newcommand{\ApJ}{Astrophys. J}
\newcommand{\PRL}{Phys. Rev. Lett.}
\newcommand{\PRD}{Phys. Rev. D}
\newcommand{\MNRAS}{Mon. Not. Roy. Astr. Soc.}

\newcommand{\PASP}{Proc. Astr. Soc. Pac.}

\newcommand{\etal}{et al.}
\newcommand{\aut}[2]{{#2.\ #1,}}
\newcommand{\laut}[2]{{#2.\ #1,}}
\newcommand{\refs}[6]{#2, #3  {#4} (#5).}

\newcommand{\mybib}[2]{\bibitem{#2}}

\newcommand{\calb}{a}
\newcommand{\rot}{\omega}
\newcommand{\poi}{p}
\newcommand{\sig}[1]{{\boldsymbol \sigma}_{#1}}

\begin{document}

\title{Benchmark Parameters for CMB Polarization Experiments}

\author{Wayne Hu$^{1,2}$, Matthew M. Hedman$^{1}$, Matias Zaldarriaga$^{3}$}
\affiliation{
{${}^{1}$}Center for Cosmological Physics, University of Chicago, Chicago IL 60637\\
{${}^{2}$}Department of Astronomy and Astrophysics and Enrico Fermi Institute, 
	University of Chicago, Chicago IL 60637\\
{${}^{3}$}Physics Department, New York University, New York NY 10003}

\address{}
\begin{abstract}
The recently detected polarization of the cosmic microwave background 
(CMB) holds the potential for revealing the physics of inflation and 
gravitationally mapping the large-scale structure of the universe, if
so called $B$-mode signals below $10^{-7}$, or tenths of a $\mu$K, can 
be reliably detected.  We provide a language for describing systematic effects 
which distort the observed CMB temperature and polarization fields 
and so contaminate the $B$-modes. We identify 7 types of effects, described
by 11 distortion fields, and show their association with known instrumental
systematics such as common mode and differential
gain fluctuations, line cross-coupling, 
pointing errors,
and differential polarized beam effects.
Because of aliasing from the small-scale structure in the CMB, 
even uncorrelated fluctuations in these effects can affect the large-scale $B$
modes relevant to gravitational waves.  Many of these problems are greatly reduced
by having an instrumental beam that resolves the primary anisotropies
(FWHM $\ll 10'$).  
To reach the ultimate goal of an inflationary energy scale of 
$3 \times 10^{15}$ GeV, polarization distortion fluctuations must be controlled
at the $10^{-2}-10^{-3}$ level and temperature leakage to the $10^{-4}-10^{-3}$
level depending on effect. For example pointing errors must be controlled to 
$1.5''$ rms for arcminute scale beams or a percent of the Gaussian beam width
for larger beams; low spatial frequency 
differential gain fluctuations or line cross-coupling 
must be eliminated at the level of $10^{-4}$ rms. 
\end{abstract}
\maketitle

The recently detected polarization of the cosmic microwave background \cite{Kovetal02} 
holds subtle imprints in its pattern that potentially can reveal the physics of 
the inflationary epoch  \cite{KamKosSte97,ZalSel97} and provide a new handle on 
the dark matter and energy in the universe \cite{ZalSel98,BenBervan00,HuOka02}.   
This curl pattern, the so-called $B$-modes,
lies at least an order of magnitude down in amplitude compared with the detected main polarization level, which itself is an order of magnitude lower than the temperature
anisotropy.  Clearly their detection represents a substantial experimental challenge.

Beyond raw sensitivity requirements for instruments \cite{JafKamWan99},
much attention has already been given in the literature to two aspects of 
this challenge:
astrophysical foregrounds (e.g. \cite{TegEisHudeO00,PruSetBou00,Bacetal02}) 
and the survey mask and pixelization (e.g. \cite{LewChaTur01,BunZalTegdeO02}).   
The general requirements imposed on experiments are clear: multiple frequency channels 
and large,  contiguous, finely pixelized areas of sky.   
The requirements on other instrumental properties has received less attention, 
in part due to the lack of a common 
language to express their effect on $B$-modes.  Such a language must be
expressed in the map, not purely instrument, domain since $B$-modes 
reflect a spatial pattern of polarization, not its state. 
In this paper, we seek to provide
such a connective language and conduct an exploratory study on the impact of 
these systematic effects on the science of $B$-modes.

We divide polarization effects into two categories: those which are associated 
with transfer between polarization states of the incoming radiation, 
mainly induced by the detector system (\S \ref{sec:transfer}), and those which 
are associated with the anisotropy of CMB polarization and temperature, 
mainly induced by the finite resolution or beam of the telescope 
(\S \ref{sec:beam}).  We evaluate their effect on $B$-mode science in
\S \ref{sec:bmodes} and on polarization statistics in general in the Appendix.

\section{Polarization Transfer}
\label{sec:transfer}

We begin by reviewing the standard transfer matrix formalism for 
describing polarization detectors in \S \ref{sec:transferdes}  
and illustrate its use in describing the 
errors in simple polarimeters in
\S \ref{sec:transferins}.  The translation to the map domain 
is discussed in \S \ref{sec:transfermap}.

\subsection{Description}
\label{sec:transferdes}

The polarization state of the radiation is described by the intensity 
matrix $\left< E_i E_j^* \right>$ where ${\bf E}$ is the electric field
vector and the brackets denote time averaging.  As a hermitian matrix, it
can be decomposed into the Pauli basis
\begin{align}
{\bf P} &= C \left< {\bf E} {\bf E}^\dagger \right>  \nonumber \\
        &= \Theta {\bf I} 
	+ Q \sig{3} 
	+ U \sig{1} 
	+ V \sig{2} 
\,,
\end{align}
where we have chosen the constant of proportionality so that the Stokes parameters
($\Theta$,$Q$,$U$,$V$) have units of temperature under the assumption of a blackbody
spectrum.  Note that the Stokes parameters are recovered from the matrix
as (tr[${\bf I}{\bf P}$]/2,\ldots,tr[$\sig{2}{\bf P}$]/2).  
We will assume that
$V=0$ on the sky.

The instrumental response
to the radiation modifies the incoming state before detection and is generally
described by a transfer or Jones matrix ${\bf J}$ (e.g. \cite{Tin96}, see \cite{ODe02} for
an introduction in the CMB context), where
\begin{equation}
{\bf E}_{\rm out} = {\bf J}\, {\bf E}_{\rm in} \,.
\end{equation}
The polarization matrix is then transformed as
\begin{equation}
{\bf P}_{\rm out} = {\bf J} {\bf P}_{\rm in} {\bf J}^{\dagger}\,.
\label{eqn:transform}
\end{equation}
With an estimate of the transfer matrix of the instrumental response $\hat {\bf J}$,
the incoming radiation can be recovered as
\begin{align}
{\hat{ \bf P}}_{\rm in} & = \hat {\bf J}^{-1} {\bf P}_{\rm out} ({\hat{\bf J}}^{\dagger})^{-1}
\nonumber\\
			& = 
( {\hat{\bf J}}^{-1} {\bf J})
{\bf P}_{\rm in} 
( {\hat{\bf J}}^{-1} {\bf J})^{\dagger}\,.
\end{align}
The {\it errors} in the transfer matrix determination will then mix the determined
Stokes parameters according to the general transformation rule (\ref{eqn:transform})
with a new transfer matrix
\begin{equation}
{\hat {\bf J}}^{-1} {\bf J} = {\bf I} + {1 \over 2}
\left(
\begin{array}{cc}
a_c + \gamma_{1c} &   \gamma_{2c} - 2 w_c \\
\gamma_{2c} + 2 w_c &   a_c - \gamma_{1c}
\end{array}
\right)\,,
\end{equation}
where we parameterized the components with a set of 4, possibly complex, numbers
($a_c$, $\gamma_{1c}$, $\gamma_{2c}$, $w_c$).

Now let us evaluate the error in the Stokes parameters to first order in the
real and imaginary parts of the error parameters (e.g. 
Re$(a_c) \equiv a$,
Im$(a_c) \equiv a_i$)
\begin{align}
\delta (Q \pm i U) & \equiv (\hat Q \pm i \hat U) - (Q \pm i U) \nonumber\\
		&= (a \pm i 2 \omega) (Q \pm i U)  
+ (\gamma_{1} \pm i \gamma_{2})\Theta \,.
\label{eqn:qpmiu}
\end{align}
The main effects are a miscalibration of the polarization amplitude 
described by $a$,
a rotation of the orientation by an angle $\omega$, and a 
``shearing'' of the temperature
signal into polarization described by $(\gamma_{1},\gamma_{2})$, which 
we will call 
monopole leakage for reasons that will be clear below.
Note that the imaginary pieces cancel to first order and do 
not appear in Eqn.~(\ref{eqn:qpmiu}).
Furthermore terms that couple the pair ($Q + i U$, $Q - i U$) arise 
only at second order,
but we shall see that this need only be true if the Stokes parameters are measured through the
same transfer system.
Note that in the CMB context, monopole leakage from $(\gamma_1,\gamma_2)$ is particularly dangerous 
since the isotropic signal is a factor of $10^6$ or more larger than the expected polarization.  

\begin{figure}
\centerline{\epsfxsize=2.5truein\epsffile{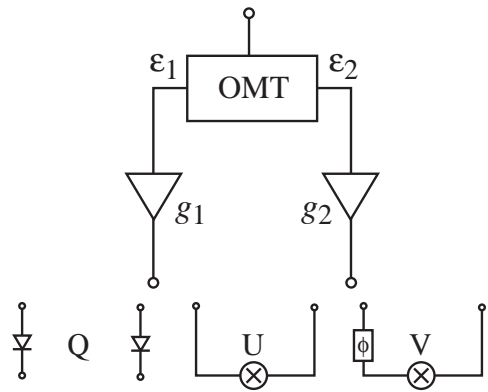}}
\caption{Block diagram for simple polarimeters.  The orthomode transducer (OMT) separates
two orthogonal linear polarization states with a leakage between the two characterized
by $(\epsilon_1,\epsilon_2)$.  After amplification with gain fluctuations $(g_1,g_2)$
the polarization state is detected by one or more of the following techniques: 
differencing the lines to produce $Q$, correlating the lines to produce $U$, correlating
the lines with a phase shift $\phi = \pi /2$ to produce $V$.  The roles of $Q$ and $V$
may be interchanged by placing a quarter-wave plate at the front end.}
\label{fig:block}
\end{figure}

\subsection{Instrumental Correspondence}
\label{sec:transferins}

Let us consider a few simple polarimeters that directly measure the polarized signal 
from a single spot 
on the sky (see Fig.~\ref{fig:block}; for
state of the art techniques, see e.g. \cite{StaChu01}).  Here the
incoming radiation is split into two, ideally orthogonal, components, $E_1$ and 
$E_2$ (using, for example an ortho-mode transducer or OMT). These components are 
possibly amplified and coupled into a detector that measures the 
polarization either
by differencing or by correlation. Ideally the transfer matrix of the components 
that split and couple the radiation into the detector is proportional to the 
identity matrix: $\hat{\bf J} \propto {\bf I}$.
 In reality it contains systematic 
errors so that ${\hat {\bf J}}^{-1} {\bf J} \propto {\bf J}$.  Let us
parameterize these errors as \cite{Heietal01}
\begin{equation}
{\hat {\bf J}}^{-1}{\bf J} = \left( 
\begin{array}{cc}
 1 + g_1 		 & \epsilon_1 e^{i \phi_1}\\
 \epsilon_2 e^{-i \phi_2} & (1 + g_2) e^{i \alpha}
\end{array}
\right)\,,
\label{eqn:jfeed}
\end{equation}
where $g_{1,2}$ are fluctuations in the gains (or, more generally, coupling 
efficiencies) of the two lines, $\alpha$ is the
phase difference between the lines, $\epsilon_{1,2}$ express the 
non-orthogonality 
or cross-coupling between the lines, and $\phi_{1,2}$ are the phases of these 
couplings.

First consider the simple differencing of the time averaged intensity 
in
the two lines $\left< E_1 E_1^* \right> - \left< E_2 E_2^* \right>$.  This forms
an estimate of $Q$ in a coordinate system attached to the instrument (e.g. \cite{Heietal01}).
Bolometer systems can be 
modeled 
with this set up, although the means of separating the two polarization states 
and the exact meaning of the parameters ($g_1$, $g_2$) 
may differ (see \cite{JonBhaBocLan02} for polarization sensitive bolometers).
Under the assumption that $g_{1,2}$, $\epsilon_{1,2}$, $\alpha \ll 1$,
\begin{align}
\delta Q &= (g_1 + g_2) Q - ( \epsilon_2 \cos\phi_2
	 - \epsilon_1 \cos \phi_1) U
\nonumber\\
	& \quad	+ (g_1 - g_2)\Theta\,,
	\end{align}
so common-mode gain fluctuations act as a normalization error $a = (g_1 + g_2)$ 
on $Q$, 
the cross-couplings act as a rotation 
$\omega =  (\epsilon_2 \cos\phi_2 - \epsilon_1 \cos\phi_1)/2$ 
and differential gain fluctuations leak temperature into polarization 
$\gamma_1 = (g_1 - g_2)$.

Now consider a simple correlation polarimeter where the signal in the two lines
are correlated as 
$\left< E_1 E_2^* \right>$ 
which forms an estimate
of $U$ in the instrument basis (e.g. \cite{Heietal01}).   
Then the errors in the determination become
\begin{align}
\delta U &= (g_1 + g_2) U + (\epsilon_2 \cos\phi_2 - \epsilon_1 \cos \phi_1)Q
\nonumber\\
         &\quad	+ (\epsilon_1 \cos\phi_1 + \epsilon_2 \cos\phi_2)\Theta\,,
\end{align}
so again $a = (g_1+g_2)$, 
$\omega = (\epsilon_2 \cos\phi_2 - \epsilon_1 \cos\phi_1)/2$ but leakage from 
$\Theta$ into $U$ is given by
$\gamma_2 = (\epsilon_1 \cos\phi_1 +
\epsilon_2 \cos\phi_2)$.  Instead of differential gain fluctuations, the 
cross-coupling between the lines is responsible for the monopole leakage in a 
correlation system.  

Notice that under the assumption of $\alpha \ll 1$ and vanishing intrinsic $V$, 
the phase error $\alpha$ does not appear to first order.  
It is instructive to consider 
the case where $\alpha$ is large, say $\alpha=\pi/2$.  In this case the correlation 
polarimeter actually measures $V$ not $U$ (see Fig.~\ref{fig:block}).  
In general, the phase error $\alpha$ rotates $U$ into $V$.

A complex correlation polarimeter actually takes advantage of the ($U$,$V$) rotation
to measure ($Q$,$U$) simultaneously (e.g. \cite{Coretal99}). 
Here, circular polarization states are coupled into the lines using,
for example a quarter wave plate before the OMT.  This effectively
converts $Q$ into $V$ in the instrument basis.
The Jones matrix of the quarter wave plate is 
\begin{equation}
{\bf J}_{1/4}(\theta)  =  {1 \over \sqrt{2}}\left(
\begin{array}{cc}
- \cos 2\theta - i  & \sin 2\theta  \\
 \sin 2\theta        & \cos 2\theta  -i 
\end{array}
\right)\,,
\end{equation}
where $\theta$ gives the orientation of the plate with 
respect to the OMT (ideally $\theta=\pi/4$). After amplification, the signal can 
be coupled into two 
different correlators, which include different phase shifts between the lines.
These additional phase shifts can be represented
with the transfer function:
\begin{equation}
{\bf J}_{\rm phase}(\phi) = \left(
\begin{array}{cc}
1 & 0 \\
0 & e^{i \phi} 
\end{array}
\right)\,.
\end{equation}
For one correlator $\phi$ is set to zero, yielding an estimate of $U$, 
while the 
other correlator has $\phi=\pi/2$, providing an estimate of $Q$. 

Now let us consider the effect of certain imperfections.  Consider the 
actual transfer matrices of the two correlations to be 
\begin{align}
{\bf J}_U &= 
	  {\bf J}_{\rm line}(g_1,g_2,\epsilon_1,\epsilon_2) 
	  {\bf J}_{1/4}(\pi/4 + \beta)\,,\\
{\bf J}_Q &= {\bf J}_{\rm phase}(\pi/2+\psi) 
	  {\bf J}_{\rm line}(g_1,g_2,\epsilon_1,\epsilon_2) 
	  {\bf J}_{1/4}(\pi/4 + \beta)\,,\nonumber
\end{align} 
and the assumed transfer matrices to be
\begin{align}
\hat {\bf J}_U &= 
	  {\bf J}_{1/4}(\pi/4)\,,\\
\hat {\bf J}_Q &= {\bf J}_{\rm phase}(\pi/2) 
	  {\bf J}_{1/4}(\pi/4)\,,\nonumber
\end{align} 
where the line matrix is taken from Eqn.~(\ref{eqn:jfeed}) with the phase factors
set to zero for simplicity.  Then the errors become
\begin{align}
\delta(Q \pm i U) &=[(g_1 + g_2) \pm 2 i \beta](Q\pm i U)  \nonumber\\
		  &\quad  + \psi\, U + (\epsilon_1 +\epsilon_2 )\Theta\,.
\end{align}   
The new feature in this system that was not present in the simple correlation 
polarimeter is an asymmetry between $Q$ and $U$ which is first
order in the phase error $\psi$.  More generally, a technique that
simultaneously measures $Q$ and $U$ may have separate transfer
properties (calibration, rotation, etc.) that appear as a coupling of
opposite spin states $Q + i U$ and $Q - i U$.  We will call such effects
spin flip terms.

The Jones matrix formalism can be applied to more complicated polarimeters, 
such as inteferometers, or other systematics such as the finite 
emissivity of the dish polarimeters \cite{Far02}.
In general, the systematic errors in the detector system will lead
to calibration errors, rotation of linear polarization, leakage of temperature
into polarization, and coupling between the two spin states $Q \pm i U$.

\subsection{Map distortions}
\label{sec:transfermap}

Errors in the polarization sensitivity of the detector system that vary with time
will translate into
errors in the polarization sky maps that vary with position.  Map making generally 
proceeds by modeling the time ordered data 
as a vector of numbers 
${\bf d}$ (e.g. \cite{deOetal02})
\begin{equation}
{\bf d} = {\bf A} {\bf s} + {\bf n}\,,
\end{equation}
where ${\bf n}$ is the instrumental noise and
${\bf s}$ is the model of the signal, say $\{ Q(\bn_1),U(\bn_1)..., 
Q(\bn_{n_{\rm p}}), U(\bn_{n_{\rm p}}) \}$ 
for a sky map with $n_{\rm p}$ pixels.  Here ${\bf A}$ is the pointing matrix and
in its simplest incarnation just encodes the sky pixel
at which the instrument is pointed at the given time. 
More generally the pointing
matrix also encodes the beam and the chopping strategy where different 
pointings are differenced to remove systematic offsets.  The
additional complication for polarization is that the pointing matrix also
has to encode the orientation of the instrument to transform
$(Q,U)$ in the instrument basis to the fixed sky. This is an advantage
since systematic errors like the monopole leakage $(\gamma_1,\gamma_2)$
are fixed to the instrument basis and not the sky.

Given the statistical properties of
the noise ${\bf N} = \langle {\bf n}{\bf n}^t \rangle$, the minimum variance 
map reconstruction is 
\begin{equation}
\hat {\bf s} =  [{\bf A}^t {\bf N} {\bf A}]^{-1} {\bf A}^t {\bf N} {\bf d}\,.
\end{equation}
This weighting of the data vector then also describes the transformation of
the instrumental systematic errors to errors in the map.  In the simplest
case of white detector noise, fixed instrument orientation, simultaneous
$Q$ and $U$ detection and no chopping,
the weighting simply averages the $n_{\rm p}$ separate pointings 
for each pixel.  If
the systematic fields, e.g. the calibration error $a(t)$ were uncorrelated in time,
they would remain so in the map but with a variance that is reduced by 
${n_{\rm p}}$.   Low frequency temporal correlations in 
the systematics will produce correlated
noise in the map.  This is generally controlled by spatially 
cross-linking the scans \cite{WriHinBen96,Janetal96}.  
A noise power of the $1/f$ form will typically lead
to spatial correlations between white and $1/l$ \cite{Teg97}, where $l$ is the angular frequency or
multipole moment (see \S \ref{sec:bfield}).
Note that even a $1/l$ spectrum gets most of its variance from 
high $l$ and so contamination at the pixel or beam scale will
be of particular interest in \S \ref{sec:bmodes}.

Since the translation between the temporal and map domain is conceptually 
straightforward
but highly dependent
on the scanning strategy, we parameterize the systematic errors directly
in the map
\begin{align}
\delta [Q \pm i U](\bn) &= 
			[\calb \pm i 2 \rot](\bn)  [Q \pm i U](\bn) \nonumber\\
			& \quad + [f_1 \pm i f_2](\bn)   [Q \mp i U](\bn) \\
		        & \quad + [\gamma_1 \pm i \gamma_2](\bn) \Theta(\bn)\,. \nonumber 
\end{align}
These correspond to calibration and rotation, spin-flip coupling and monopole
leakage errors as they appear in the map.

\section{Local Contamination}
\label{sec:beam}

In the previous section, we dealt with polarization transfer 
in a single, perfectly known, direction on the sky.  An experiment necessarily has 
finite resolution and thus there is an additional class of contamination associated with 
the resolution or beam of the experiment.  We will consider here contamination from 
a local coupling between the Stokes parameters which models 
low order anisotropy in the polarized beams. 

\subsection{Description}

Let us consider the polarization fields to be mixed locally 
\begin{align}
\label{eqn:localmodel}
\delta[Q \pm i U](\bn;\sigma) &= \sigma {\bf p}(\bn) \cdot \nabla [Q \pm i U](\bn;\sigma) \\
&\quad  +  \sigma [d_1 \pm i d_2](\bn) [\partial_1 \pm i\partial_2] \Theta(\bn;\sigma)
\nonumber\\
&\quad  + \sigma^2 q(\bn) [\partial_1 \pm i \partial_2]^2 \Theta(\bn;\sigma)\,, \nonumber
\end{align}
where the fields are smoothed over the average beam of the experiment, here denoted by
$\sigma$, the Gaussian width.  Therefore the fields ${\bf p}$, ${\bf d}$ and $q$ 
represent sensitivity to structure in the fields
on the scale of the beam.  We will call these the pointing error, dipole leakage
and quadrupole leakage respectively.

We truncate the local coupling at the dipole level for the polarization and the
quadrupole level for the temperature.  
These terms have a direct correspondence to known systematics 
as we shall see.  
More generally, the form of these couplings is dictated by the properties 
of the polarization field under rotation and can be generalized to
higher order in a straightforward manner.

\subsection{Instrumental Correspondence}
\label{sec:beaminstr}
Local couplings are primarily due to imperfections in the beams. Even a 
perfectly on-axis, azimuthally symmetric telescope will not in general 
produce a 
completely azimuthally symmetric and perfectly polarized beam. In general the 
beam has a finite ellipticity along the axis of polarization, and there is a
``cross-polar" beam which couples to the ``wrong" polarization state 
(additional asymmetries may appear in off-axis telescopes \cite{Heietal01}). 
Both these imperfections arise because the surface normal 
to the optics has different orientations with respect to polarization axis 
depending on where the incident radiation strikes the telescope 
\cite{Rum66,Tin96}. 
  
As an illustrative
model of these effects, consider the case where the receiver on the 
telescope is a simple differencing polarimeter. Here the effects of the 
cross-polar beams are of second order, and will be ignored. Radiation
 from the  sky is then coupled into one line of the detector through a perfectly 
polarized beam:
\begin{align}
B(\bn; {\bf b}, e) &= {1 \over 2\pi \sigma^2 (1-e^2)}\exp \Big[
 -{1 \over 2\sigma^2}\Big( {(n_1-b_1)^2 \over (1+e)^2}\nonumber\\
     &\quad +
{(n_2-b_2)^2 \over (1-e)^2} \Big) \Big]
		   \,,
\label{eqn:beammodel}
\end{align}
where ${\bf b}$ is the offset 
between the beam center
and the desired direction on the sky, $\sigma$ is the mean beamwidth, and $e$ 
is the ellipticity \cite{Heietal01b}.
These parameters are different for the different polarizations, and the 
difference in the beams enters into the  $Q$ measurement
\begin{equation}
B(\bn;{\bf b}_a,e_a)-B(\bn;{\bf b}_b,e_b)\,.
\end{equation}
To first order in the sums and differences of the ellipticities 
and pointing errors
\begin{align}
\sigma {\bf p} &= ({\bf b}_a + {\bf b}_b)/2 \,, \nonumber\\
\sigma {\bf b}_d &= ({\bf b}_a - {\bf b}_b)/2 \,, \nonumber\\
e_s &= (e_a + e_b)/2  \,,\nonumber\\
q &= (e_a - e_b)/2 \,, 
\end{align}
we obtain
\begin{align}
\hat Q(\bn;\sigma) &= \int d \bn' B(\bn') \Big\{ 
	Q(\bn + \bn' + \sigma {\bf p}) \nonumber\\
       & \quad  + \Big[ ({{\bf b}_d  \cdot {\bn'} \over \sigma }) 
	+ {q \over \sigma^2}({n_2'}^{2} - {n_1'}^{2}) \Big]\Theta(\bn+\bn') \Big\}\,, \nonumber\\
       & \approx Q(\bn;\sigma) +  \sigma {\bf p} \cdot \nabla Q(\bn;\sigma) +     
	\sigma {\bf b}_d \cdot \nabla \Theta(\bn;\sigma) \nonumber\\
       & \quad + \sigma^2 q [\partial_1^2 - \partial_2^2] \Theta(\bn;\sigma) \,,
\end{align}
where the average beam $B(\bn) = B(\bn;0,0)$ and we drop second derivative terms in 
$Q$.
A difference in the  mean beamwidth of the two beams has the same form as a contribution
to the monopole leakage except that the filter for the
temperature field is the beam difference not beam and is not simply a low pass filter.

A pointing offset in both beams becomes a 
gradient coupling in polarization and a
differential beam ellipticity or ``squash'' becomes a coupling to the temperature quadrupole.  These
have a clear correspondence to the local contamination model of Eqn.~(\ref{eqn:localmodel}).
A differential pointing offset, or ``squint'' translates into coupling to the 
temperature dipole.   The
exact correspondence to the model is not precise since the leakage does not truly behave
as a false polarization.  For example under rotation of the instrument by $\pi$, the
false $Q$ reverses sign.  The model of Eqn.~(\ref{eqn:localmodel}) with ${\bf d}$ referenced
to the sky (not the instrument) does transform as polarization and so should be viewed
as the residual dipole sensitivity after correction.  
The quadrupolar coupling
is particularly dangerous since it behaves precisely as a polarization and
cannot be removed through rotation of the instrument.   For example, even
a circularly symmetric temperature hot spot becomes a radial pattern of polarization
through its local quadrupole moment.   

These leakage terms also appear if the receiver is a correlation polarimeter. In 
this case, the leakage from temperature to polarization is due to the 
amplitude and shape of the cross-polar beam instead of asymmetries in the main 
beam \cite{Heietal01}. However, because both of these imperfections have a common origin in 
the variations in the boundary conditions at optical surfaces, it turns out that 
a differencing system or a correlation polarimeter will obtain similar leakage 
terms due to the local effects (after accounting for 
the orientation of the receiver with respect to the optics) \cite{Nap89,Zha93}. 
Interferometric polarimeters have related effects on the scale of the primary beam
although some of the implications for $B$-modes will differ since the measured
modes are below the beam scale \cite{Leietal02}.

If stable, these effects can be removed given a beam measurement and the
true temperature field on the sky using the formalism of
anisotropic polarized beams \cite{Chaetal00,FosDorBou02}.  Moreover, as the circularly
symmetric hot spot example implies, a stable quadrupole leakage produces no $B$-mode
in the polarization map (see \S\ref{sec:bdistortion}).
It is the instability in these effects or errors
in the subtraction that appear as errors in the map.

\section{B-mode Contamination}
\label{sec:bmodes}

We study here the implications of polarization transfer and local contamination on the $B$-modes of the polarization.  In \S \ref{sec:bfield}, we give
the harmonic representation of the polarization and contamination fields. In
\S \ref{sec:bdistortion}, we compute the contamination
to the $B$ power spectrum from polarization distortion and temperature leakage.
We explore the implications of these effects in \S \ref{sec:bimpact}.

\subsection{Field Representation}
\label{sec:bfield}

The polarization and contamination fields may in general be decomposed into harmonics
appropriate to their properties under rotation or spin.  For small sections of
the sky, these harmonics are simply plane waves \cite{Sel97,WhiCarDraHol99}; 
in the Appendix we treat the
general all-sky case.  We will follow the 
convention that a complex field $S$ of spin $\pm s$
is decomposed as
\begin{align}
[S_1 \pm i S_2](\bn) &= (\mp 1)^s \int { d^2 l \over (2\pi)^2 }  [S_a \pm i S_b](\bl) e^{\pm is \phi_l}\,,
\label{eqn:bflatsky}
\end{align}
where $\cos \phi_l = l_x/l$.  
The complex polarization $Q\pm i U$ is a spin $\pm 2$ field and we will follow the conventional
nomenclature that its harmonics are named $E \pm i B$.  This property requires
the calibration $\calb$, rotation $\rot$, and quadrupole leakage to be spin-0 fields, 
the pointing $p_1 \pm i p_2$ and dipole leakage $d_1 \pm i d_2$ to be $\pm 1$ fields,
the monopole leakage $\gamma_1 \pm i \gamma_2$ to be $\pm 2$ fields, 
and the spin-flip $f_1 \pm i f_2$ to be $\pm 4$ fields. 
For spin $\pm 1$ fields $S_a$ is the divergence-free part and $S_b$ is the curl-free part. 

\begin{figure}
\centerline{\epsfxsize=3.5in\epsffile{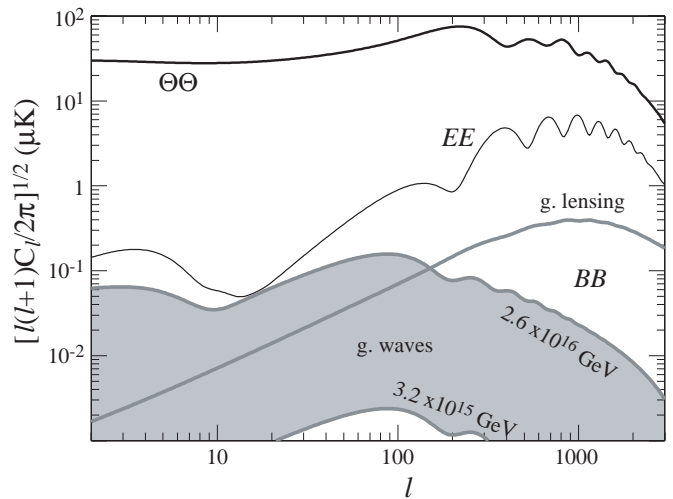}}
\caption{Scalar CMB power spectra in temperature ($\Theta\Theta$)
and $E$-mode polarization ($EE$) compared with $B$-mode polarization
due to gravitational lensing and gravitational waves at the
maximum allowable $2.6 \times 10^{16}$ GeV \cite{WanTegZal02} and minimum
detectable $3.2 \times 10^{15}$ GeV level \cite{KnoSon02}. 
The $\Lambda$CDM model shown has parameters
given in \ref{sec:bfield}.}
\label{fig:fidmod}
\end{figure}

Under the assumption of statistical isotropy of the fields, their two point correlations
are defined by their (cross) power spectra
\begin{equation}
\left< S (\bl)^*  S'(\bl') \right> = (2\pi)^2 \delta(\bl-\bl') C_l^{S S'}\,,
\end{equation}
where $S$, $S'$ are any of the fields.  In particular, the CMB polarization is
described by $C_l^{EE}$ and $C_l^{BB}$ and CMB temperature by 
$C_l^{\Theta\Theta}$.  
Note that $C_l^{BB}=0$ for scalar 
fluctuations in linear theory.  
For definiteness, let us take as a fiducial
model: a baryon density of  
$\Omega_b h^2 =0.02$,  cold dark matter density of $\Omega_c h^2 = 0.128$, 
a cosmological constant of $\Omega_\Lambda=0.65$, reionization optical depth 
$\tau=0.05$, an initial amplitude of comoving curvature fluctuations of 
$\delta_\zeta= 4.79 \times 10^{-5}$ (\cite{Hu01c} or $\sigma_8=0.92$),  
and a scalar spectral index of $n=1$
in a spatially flat universe.  
Power spectra for the fiducial model are shown
in Fig.~\ref{fig:fidmod}.  It is the large range in expected signals
that make the contamination problem for $BB$ so problematic.

We will calculate the contamination to the $B$-mode polarization power 
spectrum
assuming no intrinsic $B$-modes and generally will plot 
\begin{equation}
\Delta B \equiv \left( {l (l+1) \over 2\pi} C_l^{BB} \right)^{1/2}\,,
\end{equation}
in units of $\mu$K.
The general case is given in the Appendix.

\begin{figure*}
\centerline{\epsfxsize=6.0truein\epsffile{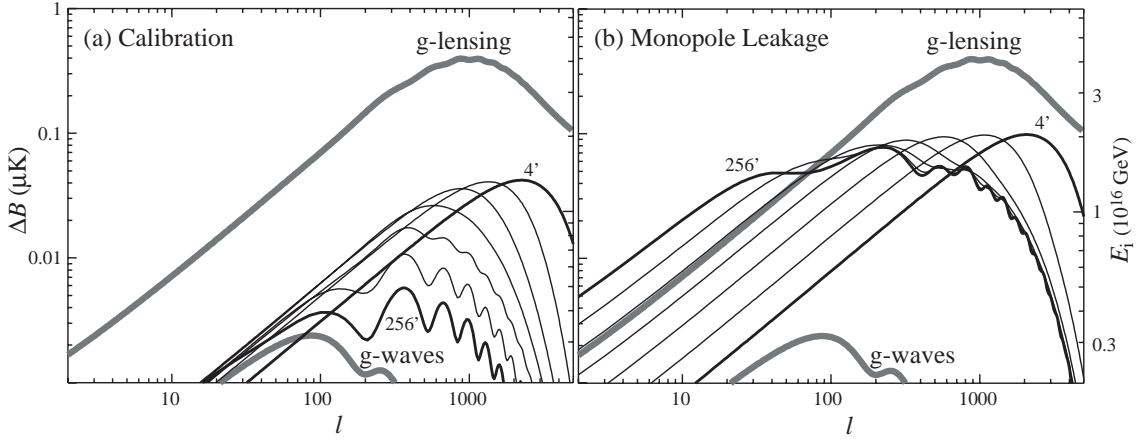}}
\caption{Coherence dependence of $B$-mode contamination
(a) for calibration $a$ with rms $A_a=10^{-2}$ 
(b) for monopole-leakage $\gamma_a$,
$\gamma_b$ with $A_{\gamma_a}=A_{\gamma_b}=10^{-3}$ added in quadrature.  
The beam scale is FWHM = $(8 \ln 2)^{1/2}\sigma = 1'$ to remove
beam effects and the FWHM coherence $(8 \ln 2)^{1/2}\alpha$ is 
stepped from $256'$ to $4'$ in factors of 2.  
Other effects follow the trend of calibration errors not monopole leakage.
For a coherence large compared with the CMB acoustic peaks, $B$ contamination
picks up their underlying structure.
Here and in the following figures, the gravitational lensing and
minimum detectable gravitational wave 
($E_i=3.2 \times 10^{15}$GeV) $B$-modes are shown for reference 
(thick shaded lines).  The scaling with $E_i$ of the peak in the $B$-mode
spectrum is shown on the right hand axis.}
\label{fig:coherence}
\end{figure*}

Although the distortion fields need not be statistically isotropic,
for illustrative purposes we will take contamination fields with power spectra
of the form
\begin{equation}
C_l^{SS} \propto \exp(-l(l+1)\alpha_{S}^2),
\end{equation}
i.e. white noise above some coherence scale $\alpha_{S}$.  The normalization
constant is set so that 
\begin{equation}
A_S^2 = \int {d^2 l \over (2\pi)^2} C_l^{SS}\,,
\end{equation}
The set ($A_S$,$\alpha_{S}$) then characterizes the rms and coherence
of the contamination field.

\subsection{$B$-modes}
\label{sec:bdistortion}

The changes to the $B$-mode harmonics due to the calibration, rotation, spin-flip 
and pointing take the form 
\begin{equation}
\delta B(\bl) = \int {d^2 l_1 \over (2\pi)^2} S(\bl_1) E(\bl_2) W_S(\bl_1,\bl_2)\,,
\label{eqn:bdistort}
\end{equation}
with $\bl_2 = \bl - \bl_1$  and
\begin{align}
W_{\calb} &= \sin[2 (\phi_{l_2} - \phi_l)]\,,  \nonumber\\
W_{\rot}  &= 2 \cos[2 (\phi_{l_2} - \phi_l)]\,, \nonumber\\
W_{p_a}   &= \sigma (\bl_2 \times \hat \bl_1)\cdot \hat{\bf z} \sin[ 2(\phi_{l_2}- \phi_l) ]\,,\nonumber\\
W_{p_b}   &= \sigma (\bl_2  \cdot \hat \bl_1) \sin[ 2(\phi_{l_2} - \phi_l)]\,, \nonumber\\
W_{f_a}   &= \sin[2 (2 \phi_{l_1} - \phi_{l_2} -\phi_{l}) ]\,,      \nonumber\\
W_{f_b}   &= \cos[2 (2 \phi_{l_1} - \phi_{l_2} -\phi_{l}) ]\,,
\label{eqn:geometricd}
\end{align}
for the various effects.  Here ${\bl_1} = l_1 \hat \bl_1$.
These relations imply contamination to the $BB$ power spectrum of
\begin{equation}
\delta C_l^{BB} = \sum_{SS'} \int {d^2 l_1 \over (2\pi)^2} C_{l_1}^{SS'} C_{l_2}^{EE}(\sigma)
 W_S^* W_{S'}\,,
\end{equation}
where 
\begin{equation}
C_{l}^{EE}(\sigma) = C_l^{EE} \exp(-l(l+1)\sigma)
\end{equation} 
is the $EE$ power spectrum smoothed over the average beam.


Similarly the change due to temperature leakage can be described by
\begin{equation}
\delta B(\bl) = \int {d^2 l_1 \over (2\pi)^2} S(\bl_1) \Theta(\bl_2) W_S(\bl_1,\bl_2)\,,
\label{eqn:bleakage}
\end{equation}
with
\begin{align}
W_{\gamma_a} &= \sin[2 (\phi_{l_1}- \phi_l)]\,, \nonumber\\
W_{\gamma_b} &= \cos[2 (\phi_{l_1}- \phi_l)]\,, \nonumber\\
W_{d_a}      &= - (l_2 \sigma) \cos[ \phi_{l_1} + \phi_{l_2} - 2 \phi_l]\,, \nonumber\\
W_{d_b}      &=   (l_2 \sigma) \sin[ \phi_{l_1} + \phi_{l_2} - 2 \phi_l]\,, \nonumber\\
W_{q}        &= - (l_2 \sigma)^2 \sin[ 2 (\phi_{l_2} - \phi_l)]\,,
\label{eqn:geometricl}
\end{align}
leading to 
\begin{equation}
\delta C_l^{BB} = \sum_{SS'} \int {d^2 l_1 \over (2\pi)^2} C_{l_1}^{SS'} C_{l_2}^{\Theta\Theta}(\sigma)
 W_S^* W_{S'}\,,
\end{equation}
for the power spectrum contamination \cite{Additional}.  

\begin{figure*}
\centerline{\epsfxsize=6.0truein\epsffile{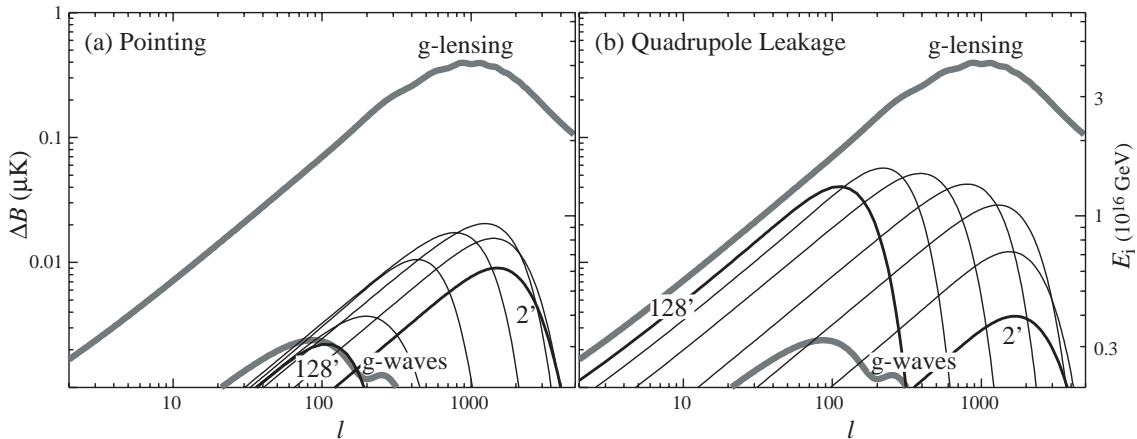}}
\caption{Beam dependence of $B$-mode contamination for 
(a) pointing with an rms $A_{p_a} = A_{p_b} = 10^{-2}$ (in units of the
Gaussian beam width) added in quadrature
(b) quadrupole leakage with an rms $A_{q}=0.002$ (in units of differential
beam ellipticity). The 
coherence $\alpha$ is set
to max($\sigma$, $10'/(8\ln 2)^{1/2}$) and the beam is stepped from $128'$ to $2'$ in
factors of 2.}
\label{fig:beam}
\end{figure*} 

A few limiting cases are worth noting before proceeding to specific examples. 
If $l_1 \gg l$ as is the case for
power in the contamination field at much smaller scales than the $l$ of interest,
$l_1 \approx l_2$ and $\phi_1 \approx -\phi_2$.  The geometric factors in 
Eqn.~(\ref{eqn:geometricd}) and (\ref{eqn:geometricl}) cause all effects to efficiently produce
$B$-modes except the pointing curl $p_a$ where the cross product vanishes.  
In the opposite limit $l_1 \ll l$ then $l_2 \approx l$ and $\phi_2 \approx \phi$.
Here the calibration $a$, pointing terms, and quadrupole leakage are geometrically suppressed.  The reason
is clear from the nature of the effects: a uniform distortion in any of these quantities does
not produce a $B$-mode. 

\subsection{Scientific Impact}
\label{sec:bimpact}

\begin{figure*}
\centerline{\epsfxsize=6.0truein\epsffile{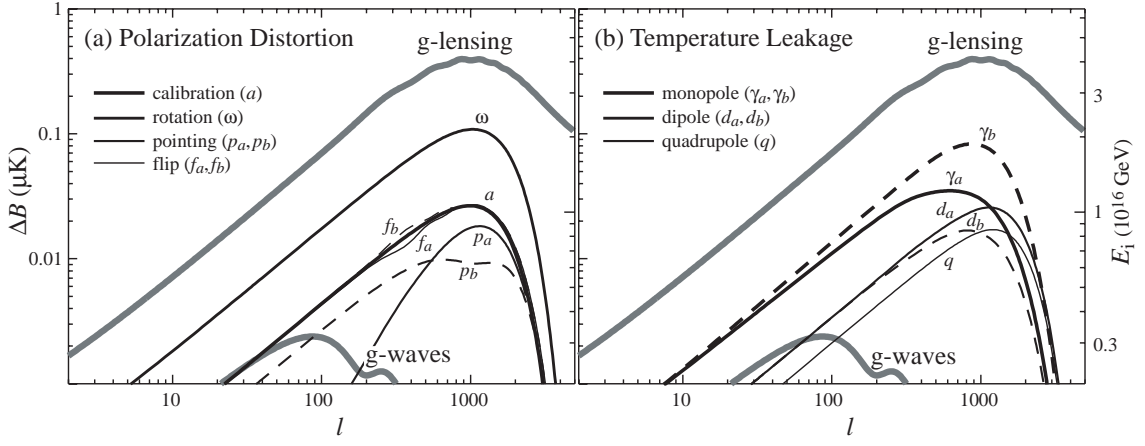}}
\caption{All effects for a beam and coherence of FWHM = $(8\ln 2)^{1/2} \sigma =10'$. 
(a) Polarization distortion for an rms
of $A=10^{-2}$ from
calibration $a$, rotation $\omega$ ($0.6^\circ$ rms), pointing ($p_a$,$p_b$) 
($2.5''$ rms) , and spin flip
($f_a$,$f_b$).  (b)  Temperature leakage for an rms of $A=10^{-3}$ from
monopole ($\gamma_a$,$\gamma_b$), dipole ($d_a$,$d_b$) and quadrupole ($q$)
terms.  The ``$b$'' component of each effect is shown with dashed lines.}
\label{fig:all}
\end{figure*}

Cosmological  $B$-modes come from two main sources: gravitational waves, also known
as tensor 
perturbations, \cite{KamKosSte97,ZalSel97} and gravitational lensing of polarization
by the large-scale structure of the universe \cite{ZalSel98}.
Aside from small but interesting effects due to 
the dark energy, reionization and massive neutrinos, 
the gravitational lensing $B$-modes can be predicted given parameters extracted
from the CMB temperature spectrum.   
The gravitational lensing prediction in the fiducial model is shown in 
Figs.~\ref{fig:fidmod}-\ref{fig:all} 
as the shaded top line.

Under slow-roll inflation, the initial amplitude of the gravitational wave spectrum is
parameterized by the energy scale of inflation $E_i$ and 
its spectrum is nearly scale invariant.  
It predicts a $B$-mode power spectrum
amplitude with a peak at $l \approx 90$ of \cite{TS}
\begin{align}
\Delta B_{\rm peak} 
&= 0.024 \left( {E_i \over 10^{16} {\rm GeV} }\right)^2 \mu{\rm K}\,.
\end{align}
Under reasonable cosmological assumptions, the CMB temperature anisotropies
constrain the energy scale to be $E_i < 2.6 \times 10^{16}$GeV \cite{WanTegZal02}.
If the energy scale is less than $E_i < 3.2 \times 10^{15}$GeV, then even with a
direct reconstruction of the lensing signal \cite{HuOka02}, 
a significant detection
of the inflationary $B$-modes cannot be achieved \cite{KnoSon02}.  
These two extremes are shown in Fig.~\ref{fig:fidmod}
and are used in Figs.~\ref{fig:coherence}-\ref{fig:all},
to mark the range across which the systematic errors
need to be controlled.  We will take the prediction for the
middle of this range ($E_i=10^{16}$ GeV)
as the minimal level that a next generation polarization mission must reduce
errors.  For reference, with no systematics or foregrounds 
the Planck satellite \cite{planck} can in 
principle achieve a 1$\sigma$ bound of $E_i = 1.1 \times 10^{16}$ GeV \cite{Hu01c}. 

Given that the inflationary $B$-modes peak at $l \approx 90$, one might naively assume that only
contamination fields with coherence corresponding to degree scales would be problematic.
However because equations (\ref{eqn:bdistort}) and (\ref{eqn:bleakage}) represent mode 
coupling, this expectation is incorrect.    The problem is that the intrinsic power in the CMB 
polarization fields as well as the temperature gradient and second derivative fields 
peak on the scale associated with 
the diffusion scale
at recombination, now observationally determined 
to be  $l \approx 10^3$ or $10'$ by the
CBI experiment \cite{Peaetal02}.    
In Fig.~\ref{fig:coherence}a, we show the effect of a
calibration error with the same rms $A_a = 10^{-2}$ but different coherence scales
$\alpha_a$.  For coherence scales above $(8\ln 2)^{1/2} \alpha_a = 10'$, the contamination
actually increases as the coherence scale decreases.  For most effects, the coherence scale 
that gives the maximum total contamination is the larger of the beam scale and $\sim 10'$. 
The mathematical reason is that the mode coupling sets $\bl = \bl_1 + \bl_2$
which forms a triangle with sides $(l,l_1,l_2)$.
For CMB power at ${\bf l}_2\gg {\bf l}$ contamination power at $l_1 
\approx l_2$ causes most of the leakage by forming a flattened triangle.  

The exception is the monopole leakage which takes power out of the 
CMB temperature power spectrum itself, not derivative power spectra which are
weighted by factors of $l$.  Here the most damaging coherence 
scale is associated with the first peak in the CMB at 
$l \approx 200$ (see Fig.~\ref{fig:fidmod}) which is 
dangerously close to the $l \sim 100$ scale of interest for gravitational waves. 
Fig.~\ref{fig:coherence}b illustrates this problem and shows that degree
scale fluctuations in monopole leakage from low frequency noise
must be controlled to substantially
better than $10^{-3}$ rms for $E_i < 10^{16}$GeV. 

Pointing, dipole and quadrupole leakage errors are expressed in terms of fractions
of the beam and hence can depend strongly on the beam scale.  The contamination
at $l \approx 90$ from pointing errors of a fixed fraction of the beam holds roughly 
constant for beam sizes above a FWHM $\approx 10'$.  At this point most of the
structure in the underlying CMB fields become resolved and the contamination depends
on the absolute pointing error relative to the CMB $10'$ coherence (see Fig.~\ref{fig:beam}).
Pointing problems must be constrained to better than the larger of
$10^{-1}$ of the Gaussian beamwidth or $15''$ absolute rms for
$E_i < 10^{16}$GeV and must reach $10^{-2}$ of the beamwidth
 or $1.5''$ absolute
to be safely irrelevant. 

The quadrupole leakage provides a more extreme example.  Contamination for
a fixed rms differential ellipticity strongly increases with increasing beam
and so a beam $< 10'$ FWHM greatly reduces the contamination.
The dipole leakage lies in between these two cases in sensitivity to the beam scale.
In Fig.~\ref{fig:all} we show all of the effects, for a choice of beam and 
coherence of FWHM $= 10'$ and an rms of $10^{-2}$ for polarization distortions
and $10^{-3}$ for temperature leakage.

It is useful to have an approximate scaling for the rms amplitude of the
systematic needed to make the contamination on the same level (at $l=90$) as 
a given target inflationary energy scale. 
Let us approximate the rms as a power law in the FWHM of the beam
\begin{equation}
A_S = C_S \left( {E_i \over 10^{16} {\rm GeV}} \right)^2 
({\rm FWHM \over 10'})^{p_S}\,.
\end{equation}
In Table 1, we give the coefficients $C_S$ and $p_S$ for two choices
of the coherence scale: $\alpha_S = \sigma$ and $(8\ln 2)^{1/2} \alpha_S = 2^\circ$.
The beam dependence is calculated locally around $10'$ and should not be used
to extrapolate results far from this.

\begin{table}[tb]\footnotesize
\begin{tabular}{llrllr}
Type & $C_S$ ($\sigma$) & $p_S$ ($\sigma$) && $C_S$ ($2^\circ$) & $p_S$ ($2^\circ$)
\\
&&&&\\
Calibration $a$ &
     0.060 &     -0.3 &\hphantom{xxx} &
     0.049 &      0.0 \\
Rotation $w$ &
     0.015 &     -0.3 &&
     0.011 &      0.0 \\
Pointing $p_a$ &
     0.75 &     -1.3 && 
     0.53 &     -1.0 \\
Pointing $p_b$ &
     0.098 &     -0.7 &&
     0.57 &      -1.0 \\
Flip $f_a$ &
     0.061 &     -0.3 &&
     0.046 &      0.0 \\
Flip $f_b$ &
     0.059 &     -0.3 &&
     0.045 &      0.0 \\
Monopole $\gamma_a$ &
     0.0023 &     -0.9 &&
     0.0006 &      0.0 \\
Monopole $\gamma_b$ &
     0.0019 &     -0.9 &&
     0.0005 &      0.0 \\
Dipole $d_a$ &
     0.0077 &     -1.3 &&
     0.0053 &     -1.0 \\
Dipole $d_b$ &
     0.0077 &     -1.3 &&
     0.0056 &     -1.0 \\
Quadrupole $q$ &
     0.0124 &     -1.5 &&
     0.0394 &     -2.0 \\
\end{tabular}
\caption{Scaling parameters for contamination effects with a coherence
of the beam scale $\sigma$ and $2^\circ$. $C_S$ represents the
minimum rms required to not exceed a signal at $E_i = 10^{16}$ GeV;
the ultimate limit of $3.2 \times 10^{15}$ GeV would require an order
of magnitude smaller rms.}
\label{tab:scaling}
\end{table}

\section{Discussion}

We have provided a fairly general description of the phenomenology of
systematic errors that can occur in polarization maps, their correspondence with
known classes of instrumental problems, and their impact on the
science of $B$-modes.    Instability in the systematic effects or errors
in their removal lead to residual contamination in the polarization maps 
that are parameterized by 7 fields, 4 of which have
two components each, for a total of 11 distortion parameters per
position on the sky or multipole moment.  These errors are 
associated with calibration, rotation, pointing (2), spin flip (2),
monopole leakage (2), dipole leakage (2) and quadrupole leakage.
The three temperature leakage effects are named for the type of temperature
fluctuation across the beam scale that they respond to and
are especially dangerous due to the
extremely low level of polarization expected in the $B$-modes. 
Monopole leakage generally arises in the receiver;
dipole and quadrupole leakage are associated with asymmetries in
the beam.

We have illustrated these problems by modelling the fluctuations
in these contamination fields with an rms amplitude and coherence.
In general, it is {\it not} sufficient to control the fluctuations in
the field on the degree scales of interest for gravitational wave
$B$-modes.   Because all of these effects transfer power from
the CMB fluctuations themselves, the most dangerous fluctuations
are those that are on the same scale as most of the power
in the CMB fields.   For all but the monopole leakage effect,
which can draw power out of the first acoustic peak, the underlying
power lies at the diffusion damping scale of $l \sim 10^3$ or 
$\sim 10'$.  Unless the beam resolves this scale, even uncorrelated
white noise fluctuations in the fields can substantially
contaminate low multipoles in $B$.  

The interplay between the beam scale and $10'$ coherence scale
of the CMB fields plays an especially important role in pointing, dipole
leakage and quadrupole leakage.  These problems couple local derivatives
of the CMB fields into false polarization signals.  They can largely
be eliminated if the beam is sufficiently small so that the CMB fields
are smooth across the beam scale.   Small beams are also desirable 
for constructing weak lensing mass maps from the $B$-mode polarization
\cite{HuOka02}.

Based on the systematic errors of the current generation of experiments,
these problems should be challenging but not insurmountable.  The DASI
instrument had percent level monopole leakage which was stable at the
fractional percent level and quadrupole leakage also at the percent level
which was highly stable.
It also had rotational uncertainties
at the percent level \cite{Leietal02}.  The PIQUE instrument 
had a monople leakage at under the percent level and a 
dipole leakage at less than 2.5\% \cite{Hed02}.  
The polarization sensitive bolometers for the upcoming Boomerang experiment
\cite{JonBhaBocLan02} and planned for Planck have monopole leakage at the percent
level but are claimed to be very stable.

This exploratory study of polarization effects should help to provide some 
rough guidance on the long
road ahead toward the ultimate goal of detecting the gravitational waves
from inflation.

{\it Acknowledgments:} 
We thank Stephan Meyer for guidance; Scott Dodelson, John Kovac, Clem Pryke, 
Bruce Winstein and the CfCP CMB Working Group for many stimulating
discussions.  WH is supported by NASA NAG5-10840 and the DOE OJI program;
MMH by the CfCP under NSF PHY 0114422, and MZ by NSF grants 
AST 0098606 and PHY 0116590
and by the David and Lucille Packard Foundation Fellowship for
Science and Engineering.
\appendix

\section{General Treatment}

The flat sky expansion in Eqn.~(\ref{eqn:bflatsky}) may be generalized by decomposing the
fields as \cite{ZalSel98}
\begin{align}
[S_1 \pm i S_2](\bn) &= (i)^s \sum_{lm} [S_a \pm i S_b]_{lm} \sYlm{\pm s}{{}}(\bn) 
\,,
\nonumber
\end{align}
where $\sYlm{s}{{}}$ is the spin-$s$ spherical harmonic \cite{spin}.

The corrections to the $E$ and $B$ harmonics of the polarization from
the distortion field $S$ can be generally expressed as 
\begin{align}
\delta X_{lm}^{\pm} &= (-1)^m \sum_{l_1 m_1}\sum_{l_2 m_2} 
	\sqrt{(2l+1)(2l_1+1)(2l_2+1) \over 4\pi} \\
     & \quad \times \wj{l}{l_1}{l_2}{-m}{m_1}{m_2} 
	S_{l_1 m_1} L_S  \nonumber\\
&\quad \times a_S^{\pm} (e_S^+ X_{l_2 m_2}^{\pm} + e_S^- X_{l_2 m_2}^{\mp})\,,\nonumber
\end{align}
where $X^+ \equiv E$, $X^- \equiv i B$
and $S \in \calb, \rot, \poi_a, \poi_b, f_a, f_b$ 
and
\begin{align}
e^\pm_{\calb,\poi_b}  &= {1 \over 2} \left[  1\pm (-1)^{l+l_1+l_2} \right],
\nonumber\\
e^\pm_{\rot,\poi_a}   &= {1 \over 2} \left[  1\mp (-1)^{l+l_1+l_2} \right],
\nonumber\\
e^\pm_{f_a}   &= \pm {1 \over 2} \left[  1\pm (-1)^{l+l_1+l_2} \right],
\nonumber\\
e^\pm_{f_b}   &= \pm {1 \over 2} \left[  1\mp (-1)^{l+l_1+l_2} \right],
\end{align}
selects out even and odd sums of the $l$'s.  The factor $a_S^{\pm} = 1$ for $S \in \calb,
\rot,\poi_a,\poi_b$ and $=\pm 1$ for $S \in f_a, f_b$ and adjusts the relative sign.

The specific linear source terms are
\begin{align}
L_\calb &= -{i \over 2} L_\rot = \wj{l}{l_1}{l_2}{-2}{0}{2}\,,\nonumber\\
L_{\poi_b} &=  {1 \over 2}
\Big[ \sqrt{(l_2+2)(l_2-1)} \wj{l}{l_1}{l_2}{-2}{1}{1}\nonumber\\
&\quad	+ \sqrt{(l_2-2)(l_2+3)} \wj{l}{l_1}{l_2}{-2}{-1}{3} \Big] \,,\nonumber\\
L_{\poi_a} &=  {-i \over 2}
\Big[ \sqrt{(l_2+2)(l_2-1)} \wj{l}{l_1}{l_2}{-2}{1}{1}\nonumber\\
&\quad	- \sqrt{(l_2-2)(l_2+3)} \wj{l}{l_1}{l_2}{-2}{-1}{3} \Big] \,,\nonumber\\
L_{f_a} &= -i L_{f_b} = \wj{l}{l_1}{l_2}{-2}{4}{-2}\,.
\end{align}

Under the assumption of statistical isotropy of the distortion fields, the
perturbation to the power spectra are given by
\begin{align}
\delta  C_l^{EE}&= 
		{a_{00} \over \sqrt{4\pi}} C_l^{EE} 
	+ \sum_{l_1 l_2 S S'} {(2 l_1 +1)(2l_2+1) \over 4\pi} C_{l_1}^{SS'} 
\nonumber\\ &\quad 
	\times	
	a_S^+ a_{S'}^+  
[ C_{l_2}^{EE} e^+_S e^+_{S'} + C_{l_2}^{BB} 
		 		e^{-}_S e^-_{S'} ] L_S^* L_{S'} \,,\nonumber\\
\delta  C_l^{BB}  &= 
		{a_{00} \over \sqrt{4\pi}} C_l^{BB} 
	+	\sum_{l_1 l_2 S S'} {(2 l_1 +1)(2l_2+1) \over 4\pi} C_{l_1}^{SS'} 
	\nonumber\\ &\quad \times 
a_S^{-} a_{S'}^{-} [ C_{l_2}^{EE} e^-_S e^-_{S'} + C_{l_2}^{BB} 
		 		e^{+}_S e^+_{S'} ] L_S^* L_{S'} \,,\nonumber\\
\delta  C_l^{EB} &= 
		{2 \omega_{00} \over \sqrt{4\pi}} (C_l^{EE}-C_l^{BB}) \\ &\quad + 
		\sum_{l_1 l_2 S S'} {(2 l_1+1)(2l_2+1) \over 4\pi} 
		C_{l_1}^{SS'} 
\nonumber\\
&\quad 
	\times a_S^{+} a_{S'}^{-} 
[ C_{l_2}^{EE} e^+_S e^-_{S'} + C_{l_2}^{BB} e^{-}_S e^+_{S'}]
		(-i) L_S^* L_{S'}\,,\nonumber
\end{align}
where we have allowed for the possibility of a monopole term in the calibration $a$
and rotation $\omega$.  Such terms often arise from second order terms in an expansion
and comes about through the variance of a distortion 
field across the sky.  They must be kept since the net change to the power spectrum
is itself second order and often cancel the linear effects.  
For example, second order terms in the pointing errors
appear as a monopole calibration error.  Since these terms do not transfer power
in $EE$ to $BB$, they are not relevant for the discussion in the main paper.

Temperature leakage terms may similarly be described in their effect on $E$ and
$iB$ 
\begin{align}
\delta X_{lm}^{\pm} & = (-1)^m \sum_{l_1 m_1}\sum_{l_2 m_2} 
	\sqrt{(2l+1)(2l_1+1)(2l_2+1) \over 4\pi} \\
      & \quad \times \wj{l}{l_1}{l_2}{-m}{m_1}{m_2} 
	S_{l_1 m_1} \Theta_{l_2 m_2} e_S^{\pm} L_S \,, \nonumber
\end{align}
for $S \in \gamma_a,\gamma_b, d_a, d_b, q$
\begin{align}
e^\pm_{\gamma_a,d_b,q}  &= {1 \over 2} \left[  1\pm (-1)^{l+l_1+l_2} \right],\nonumber\\
e^\pm_{\gamma_b,d_a}    &= {1 \over 2} \left[  1\mp (-1)^{l+l_1+l_2} \right],
\end{align} 
and 
\begin{align}
L_{\gamma_a} &= -i L_{\gamma_b} = 
	\wj{l}{l_1}{l_2}{-2}{2}{0} \,, \nonumber\\
L_{d_b} &=  i L_{d_a} = 
	\sqrt{l_2(l_2+1)} \sigma  \wj{l}{l_1}{l_2}{-2}{1}{1} \,,\\
L_{q} &= - \sqrt{ (l+2)! \over (l-2)! }\sigma^2 
	  \wj{l}{l_1}{l_2}{-2}{2}{0} \,. \nonumber
\end{align}
The perturbation to the power spectra are given by
\begin{align}
\delta C_l^{EE}  &= \sum_{l_1 l_2 S S'} {(2 l_1 +1)(2l_2+1) \over 4\pi}
		C_{l_1}^{SS'} C_{l_2}^{\Theta\Theta} \nonumber\\
&\quad e^+_S e^+_{S'} L_S^* L_{S'} \,, \nonumber\\
\delta C_l^{BB}  &= \sum_{l_1 l_2 S S'} {(2 l_1 +1)(2l_2+1) \over 4\pi}
		C_{l_1}^{SS'} C_{l_2}^{\Theta\Theta}\\
&\quad e^-_S e^-_{S'} L_S^* L_{S'} \,, \nonumber\\
\delta C_l^{EB}
&= \sum_{l_1 l_2 S S'} {(2 l_1+1)(2l_2+1) \over 4\pi} 
		C_{l_1}^{SS'} C_{l_2}^{\Theta\Theta}\nonumber\\
&\quad e^+_S e^-_{S'} (-i) L_S^* L_{S'} \,. \nonumber
\end{align}
This completes the general description of the polarization contamination from
the class of map distortions considered.

\end{document}